\documentclass[aps,prl,floatfix,amssymb,twocolumn,tightenlines,showpacs]{revtex4}

\usepackage{amsmath}
\usepackage{amsfonts}
\usepackage{latexsym}
\usepackage{graphicx}
\usepackage{color}
\usepackage{epstopdf}
\usepackage{color}

\begin{document}

\title{Demonstration of a simple entangling optical gate and its use in Bell-state analysis}
\author{N. K. Langford$^{1,2}$, T. J. Weinhold$^{1,2}$, R. Prevedel$^{1,2,3}$, K. J. Resch$^{2}$, A. Gilchrist$^{1,2}$, J. L. O'Brien$^{1,2}$, G. J. Pryde$^{1,2}$ and A. G. White$^{1,2}$}
\affiliation{$^1$Centre for Quantum Computer Technology, $^2$Department of Physics University of Queensland, Brisbane QLD 4072, Australia\\
$^3$Institut f\"{u}r Experimentalphysik, Universit\"{a}t Wien, Boltzmanngasse 5, 1090 Vienna, Austria}
\date{\today}

\begin{abstract}
We demonstrate a new architecture for an optical entangling gate that is significantly
simpler than previous realisations, using partially-polarising beamsplitters so that only a \emph{single} optical mode-matching condition is required. We demonstrate operation of a controlled-\textsc{z} gate in both continuous-wave and pulsed regimes of operation, fully characterising it in each case using quantum process tomography. We also demonstrate a fully-resolving, nondeterministic optical Bell-state analyser based on this controlled-\textsc{z} gate. This new architecture is ideally suited to guided optics implementations of optical gates.
\end{abstract}
\pacs{03.67.Lx,03.67.Mn, 03.65.Wj, 42.50.Dv \vspace{-0.31 cm}}
\maketitle

A key resource for using entanglement in quantum information protocols are gates that are capable of entangling or disentangling qubits \cite{nielsen}. Entangling gates lie at the heart of quantum computation protocols, for example, and disentangling gates used in Bell state analysers are required for quantum teleportation. Conceptually, the simplest such two-qubit gate is the controlled-\textsc{z} (\textsc{cz}) gate, which in the logical basis produces a $\pi$ phase shift on the $|11\rangle$ term, (i.e., $|00\rangle \rightarrow  |00\rangle$; $|01\rangle  \rightarrow  |01\rangle$; $|10\rangle  \rightarrow  |10\rangle$; $|11\rangle  \rightarrow  -|11\rangle$). This is a maximally entangling gate which, when coupled with single qubit rotations, is universal for quantum computing \cite{note2}.
 
In 2001, Knill, Laflamme and Milburn proposed a scheme for linear optical quantum computing which used measurement to non-deterministically realize the optical non-linearity required for two qubit entangling gates  \cite{kn-nat-409-46}. They also showed that deterministic versions of these gates could be achieved using teleportation \cite {go-nat-402-390}, which requires Bell-state measurement. Since then, there have been a number of demonstrations of quantum logic gates derived from this concept \cite{pi-pra-68-032316,ob-nat-426-264,ob-prl-93-080502,ga-prl-93-020504,ZhaoPRL}, and further theoretical development of linear optics schemes \cite{ss1,ss2,ss3,ra-pra-65-062324,ho-pra-66-024308}. In particular, there is a recent suggestion to use non-deterministic \textsc{cz} gates to construct cluster-states for demonstrating optical cluster-state quantum computation \cite{ni-prl-93-040503}.

Here we report an experimental demonstration of a non-deterministic linear-optics \textsc{cz} gate, and its application as a Bell-state analyser. This \textsc{cz} gate is the simplest entangling (or dis-entangling) linear optics gate realized to date, requiring only three partially-polarising beam splitters (PPBSs), two half-wave plates, no classical interferometers, and no ancilla photons. It is non-deterministic and success is heralded by detection of a single photon in each of the outputs. We demonstrate the operation of this type of gate using photons generated both by continuous-wave (CW) and by femtosecond pulsed parametric downconversion---by comparing the two we find that temporal mode mismatch was not a significant factor in the gate's performance. We fully characterize the operation in both regimes using quantum process tomography, and we also demonstrate the use of this kind of gate for fully-resolving Bell measurements. This simple entangling optical gate is promising for micro-optics or guided optics implementations where extremely good non-classical interference is realisable.

\begin{figure}[b]
\vspace{-0.5 cm}
\begin{center}
\includegraphics[width=8.5cm]{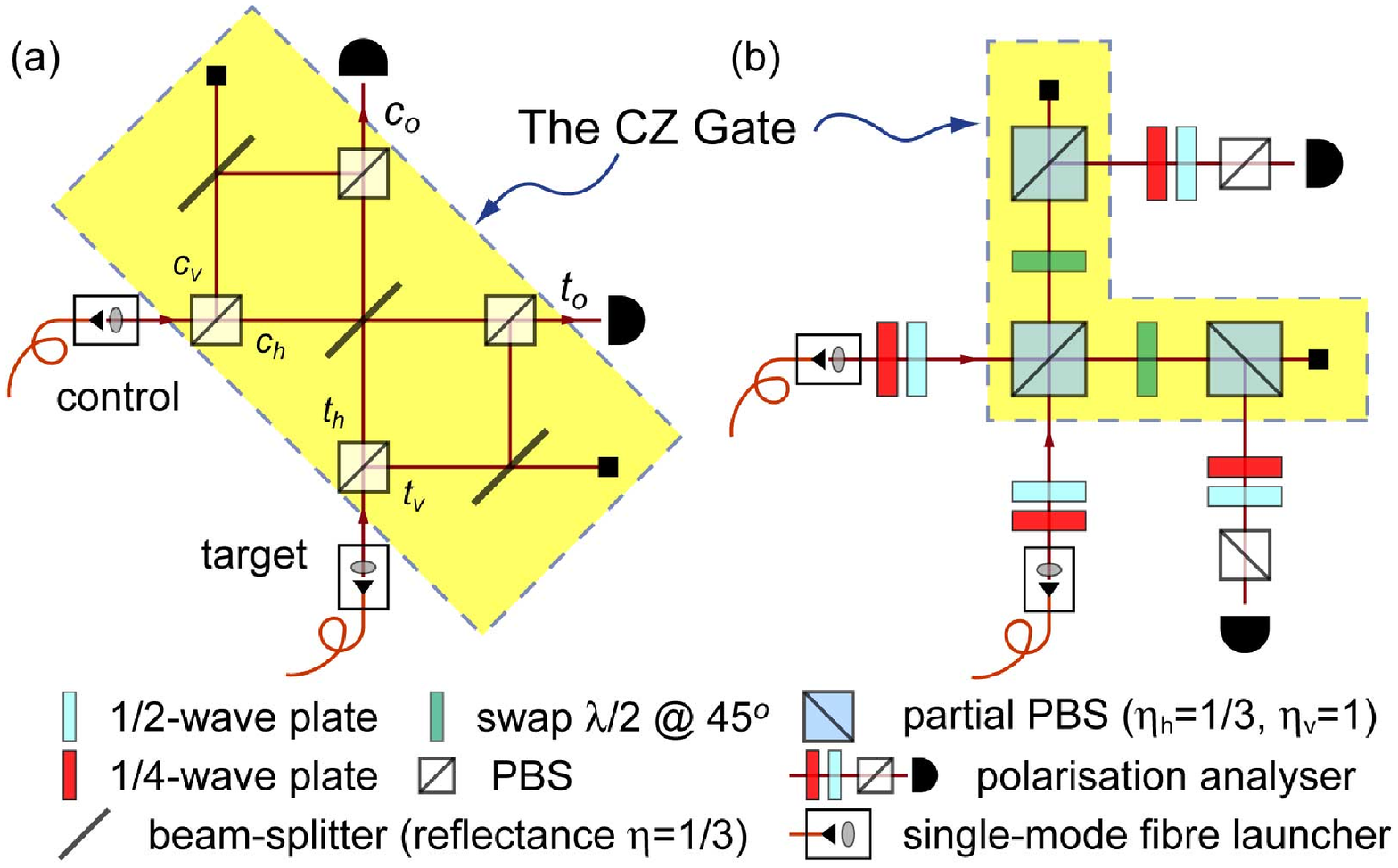}
\caption{(a) Interferometric \textsc{cz} gate based on the approach of Refs.~\cite{ra-pra-65-062324,ho-pra-66-024308}. Gate operation is enabled by transforming each qubit from polarisation to spatial encoding, and back again. This requires high interferometric stability and spatio-temporal mode-matching for correct operation. (b) Partially-polarising beam splitter (PPBS) gate. The qubits can remain polarisation-encoded, since the vertically-polarised modes are completely reflected by the first PPBS, and do not interact. Nonclassical interference occurs between the horizontally-polarised modes, with $\eta$=1/3. The subsequent PPBSs give the required losses in the $c_{V}$ and $t_{V}$ modes as shown in (a).}
\label{schematic}
\end{center}
\vspace{-0.85cm}
\end{figure}

\begin{figure*}
\centerline{\mbox{\includegraphics[width=0.66 \columnwidth]{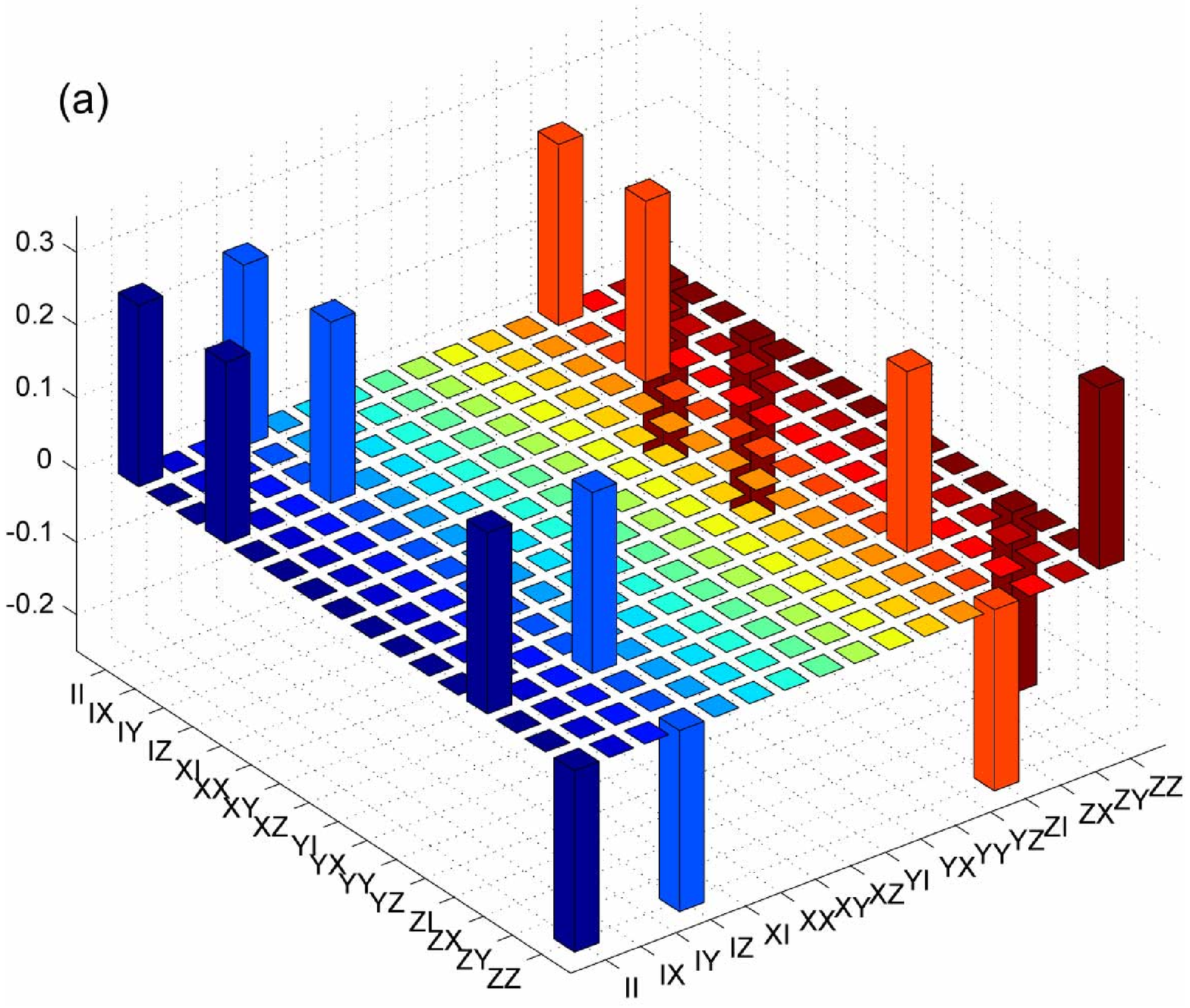}}
    \mbox{\includegraphics[width=0.66 \columnwidth]{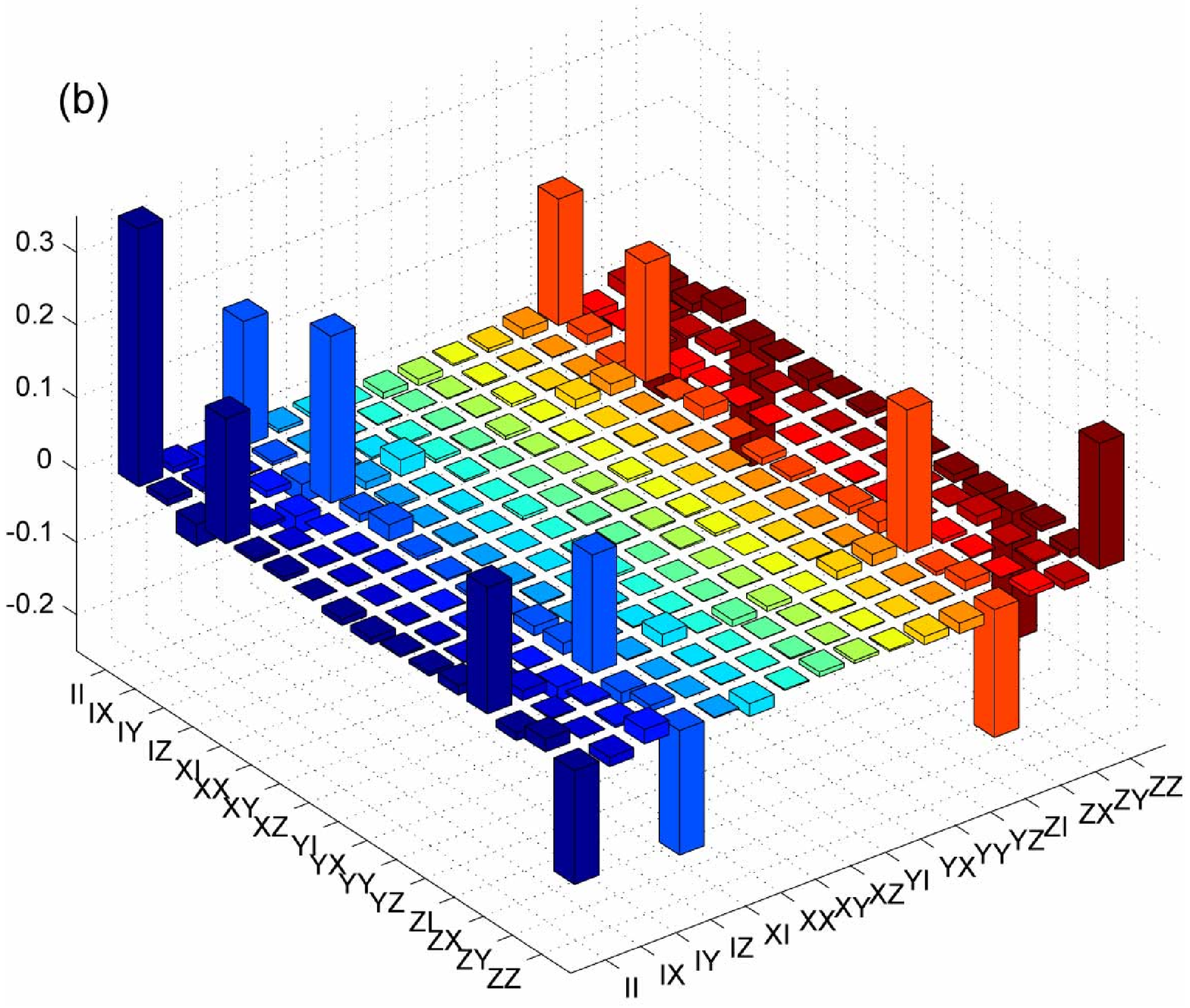}}
    \mbox{\includegraphics[width=0.66 \columnwidth]{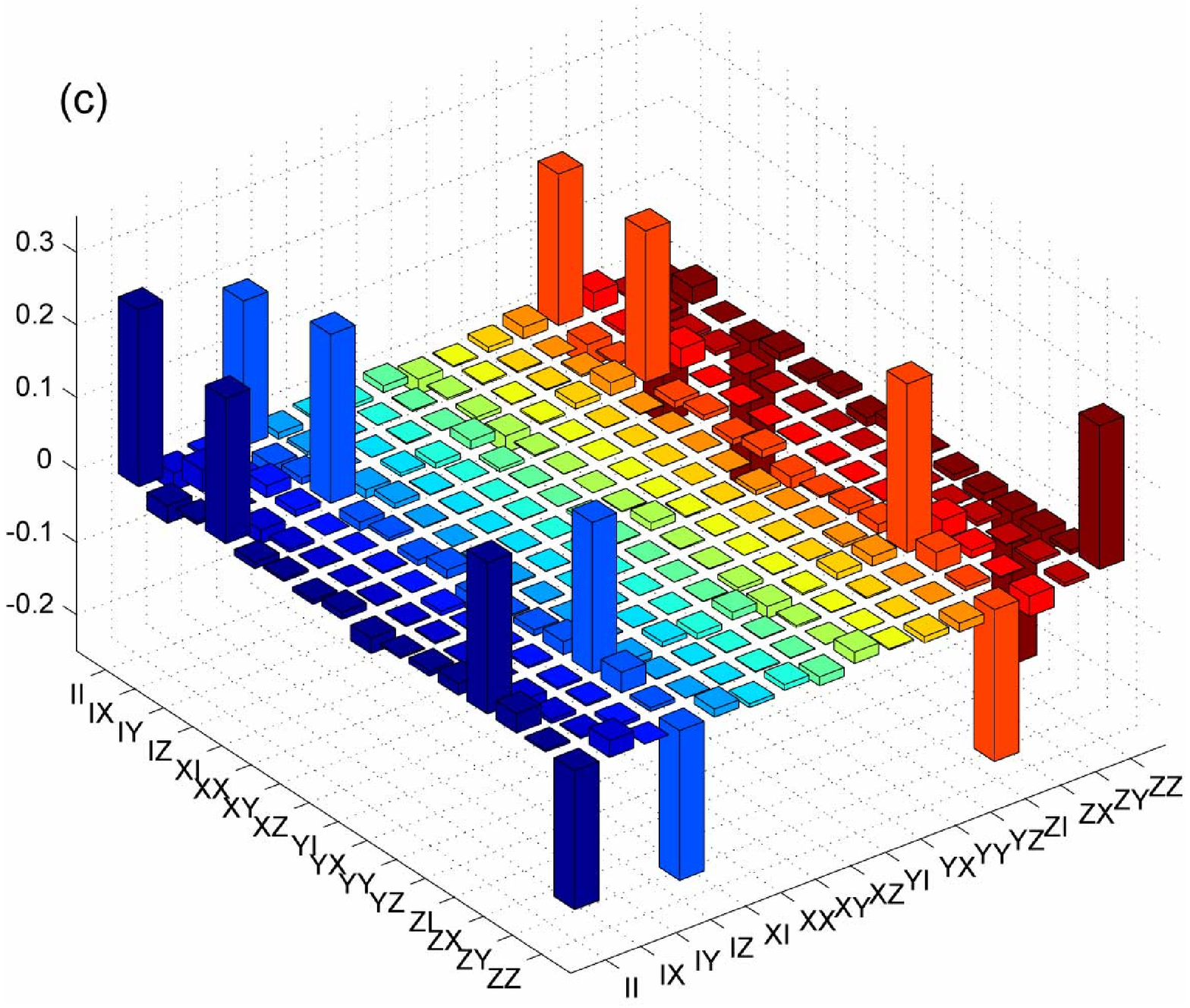}}}
\caption{Quantum process tomography of the \textsc{cz} gate. Real components of the $\chi$ matrix for the: (a) ideal; (b) CW; and (c) pulsed \textsc{cz} gate. The imaginary components of the experimental matrices are not shown: a few elements are on the order of 0.05, the average is $\sim$0.005.}
\label{qpt}
\vspace{-0.6 cm}
\end{figure*}

The best performing entangling gate implementations to date have been interferometric: a conceptual schematic of an interferometric optical \textsc{cz} gate, composed of three partially-reflecting beam splitters with reflectivity $\eta$=1/3, is shown in Fig.~\ref{schematic}(a). Each polarisation qubit input to the gate is split into two longitudinal spatial modes via a polarising beam splitter. The horizontally-polarised modes meet at a 1/3 beamsplitter and non-classical interference means that for an arbitrary input state, the entire circuit performs the transformation: $\alpha|HH\rangle+\beta|HV\rangle+\gamma|VH\rangle+\delta |VV\rangle \rightarrow \frac{1}{3}[-\alpha|HH\rangle+\beta|HV\rangle+\gamma|VH\rangle+\delta|VV\rangle]+$..., where $H$ and $V$ refer to horizontal and vertical polarisation, and the terms not shown correspond to the failure mode of the gate (i.e., the control and target output ports do not each contain one photon). With probability 1/9 the circuit performs the \textsc{cz} operation (using the logic-basis definitions, $0\equiv V$ \& $1\equiv H$). After the network of 1/3 beamsplitters, the two spatial modes of the control and target must be recombined to return to polarisation encoded qubits. Since the phase relationship between the logical modes must be maintained throughout this operation, interferometric stability is required between the control and target modes. Inherently stable interferometers have previously been used \cite{ob-nat-426-264,ob-prl-93-080502} to achieve this---however these may not be suitable for scaling to large numbers in micro- or integrated-optical realisations. Here we take an alternative approach which does not require interferometric stability.

The experimental setup for the \textsc{cz} gate we have developed is shown schematically in Fig.~\ref{schematic}(b). It employs partially polarising beamsplitters (PPBSs) that completely reflect vertically-polarized light, and have a reflectivity of 1/3 for horizontally-polarized light \cite{PPBSnote}. As in Fig.~\ref{schematic}(a), only the $H$ modes interfere nonclassically at the first PPBS. The $V$ modes each experience a 1/3 loss at the other two PPBSs: the half-wave plates between the first and subsequent PPBSs rotate polarisation by 90$^{\circ}$---a bit flip, or single qubit \textsc{x} gate---and effectively reverse the operation of the second two PPBSs. The circuit of Fig.~\ref{schematic}(b) therefore performs a \textsc{cz} gate with additional \textsc{x} gates on the control and target qubits. These additional \textsc{x} gates could be corrected by adding appropriate half-wave plates in the outputs, or by relabelling the logical states of the outputs---here we chose to relabel. The key advantage of the PPBS gate is that the polarisation modes are never spatially separated and recombined, and consequently no classical interference conditions are required. A single nonclassical interference at the first PPBS is therefore the gate's sole mode-matching condition. 

\begin{table}[b]
\vspace{-0.6cm}
  \caption{Photon source parameters}\label{dc}
  \centering 
\begin{tabular}{ccc}
\hline
\hline
Parameter   & CW  &  pulsed\\
   \hline
   \hline
Pump source   & Ar$^{+}$ & doubled Ti:Sa \\
Pump wavelength  & 351.1 nm & 410 nm \\
Crystal arrangement & Type I sandwich \cite{kw-pra-60-773} & Type I single \\
Photon wavelength & 702.2 nm & 820 nm \\
Interference filters & $\pm 0.18$ nm & $\pm 1.5$ nm\\
Output state & separable$\leftrightarrow$entangled & separable \\
\hline
\end{tabular}
\end{table}  

To test multi-qubit circuits, multi-photon sources are required. The current gold standard for generating two or more photons is pulsed parametric downconversion: pump power densities far greater than those possible with CW sources lead to significantly higher probabilities of multi-photon events. However, the short pump pulses lead to a larger bandwidth of downconverted photons, and to more stringent temporal mode-matching requirements. Thus any new gate architecture should be shown to be compatible with both CW and pulsed sources. We tested the PPBS architecture with both CW and femtosecond-pulsed sources, which produce pairs of energy degenerate single photons via spontaneous parametric downconversion in a $\beta$-Barium-Borate (BBO) crystal (Table \ref{dc}). The photon pairs were collected into single mode optical fibers to improve the spatial mode, and injected into the \textsc{cz} gate [Fig.~\ref{schematic}(b)].  In the pulsed case, mode-matching was also improved by collecting the gate output into single mode fibers.  A pair of half- and quarter-wave plates at the output of each fiber was used for input state preparation. A coincidence window of $\sim$5~ns was used and no correction for accidental counts was made. The gates were completely characterised via quantum process tomography \cite{ob-prl-93-080502, note}.  
 
A convenient representation of the measured process is the $\chi$ matrix, which is a complete and unique description of the process relative to a given basis of operators. The $\chi$ matrix for ideal \textsc{cz} gate operation in the Pauli basis is shown in Fig.~\ref{qpt}(a) (all the components are real). The experimental results for the CW gate are shown in Fig.~\ref{qpt}(b), those for the pulsed gate in Fig.~\ref{qpt}(c).  By using the method of Ref.~\cite{ob-prl-93-080502}, we are guaranteed physical $\chi$ matrices requiring no extra normalisation.  In the pulsed case, the form is similar to the ideal. In the cw case, the major deviation from ideal operation is the larger than expected \textsc{ii} term (0.36 instead of 0.25). This arises from imperfect mode matching at the first PPBS, resulting in imperfect nonclassical interference---effectively, the control and target qubits do not interfere and are simply transmitted through the circuit unchanged.

Gate performances can be quantified by calculating the process fidelity $F_P=Tr[\chi_\textrm{meas}\hspace{1pt}\chi_\textrm{ideal}]$ or the average gate fidelity,  which is the fidelity between expected and actual output states, averaged over all pure inputs, $\overline{F}=(4F_P+1)/5$  \cite{ob-prl-93-080502,gi-quant-ph-0408063}. The CW and pulsed gates have process fidelities of 74.6$\pm$0.3\%  and 84.0$\pm$0.1\% respectively; and average gate fidelities of 79.7$\pm$0.2\% and 87.2$\pm$0.1\% \cite{MCnote}. Despite more stringent temporal mode-matching requirements in the pulsed regime, the extra spatial filtering led to better gate operation, equivalent to the previous best demonstration \cite{ob-prl-93-080502}.

In our experiment, we observed systematic, fixed rotations of the input and output polarisations, probably due to birefringent effects in non-ideal PPBSs.  In practice, these have no effect on gate quality and, if necessary, could be compensated for with appropriate wave plates.  To demonstrate this, we modelled their effect numerically, identifying single qubit unitary corrections which increased the CW and pulsed process fidelities to 77.0$\pm$0.3\% and 86.6$\pm$0.2\% respectively; and average gate fidelities to 81.6$\pm$0.2\% and 89.3$\pm$0.1\%.

A potential drawback of the PPBS architecture is that the beamsplitting ratios are fixed at manufacture---in contrast to schemes where the setting of a HWP determines the effective beamsplitter reflectivity \cite{ob-nat-426-264,ob-prl-93-080502}. While the PPBSs for the CW gate (optimized for 702.2 nm) were measured to be within $\pm0.01$ of the required reflectivities, for the pulsed gate (820 nm), the values for the three PPBSs were $\eta$~=~0.28, 0.28 and 0.29 ($\pm$0.01; normalised to output power). Modelling a gate using 0.28 reflectivities, we find the optimum process fidelity that can be obtained is $F^{0.28}_P = 96$\%---near ideal. As originally anticipated in Ref.~\cite{ra-pra-65-062324}, the \textsc{cz} gate is relatively forgiving of the exact splitting ratios---making it an eminently suitable gate to be realised with a PPBS architecture. Performance of the PPBS gates are limited almost exclusively by mode matching, primarily spatial, making these gates promising candidates for micro-or integrated-optical implementations, where nonclassical mode matching in excess of 99\% can be expected \cite{pi-prl-90-240401}.

\begin{figure}[t]
\begin{center}
\includegraphics[width=8.5cm]{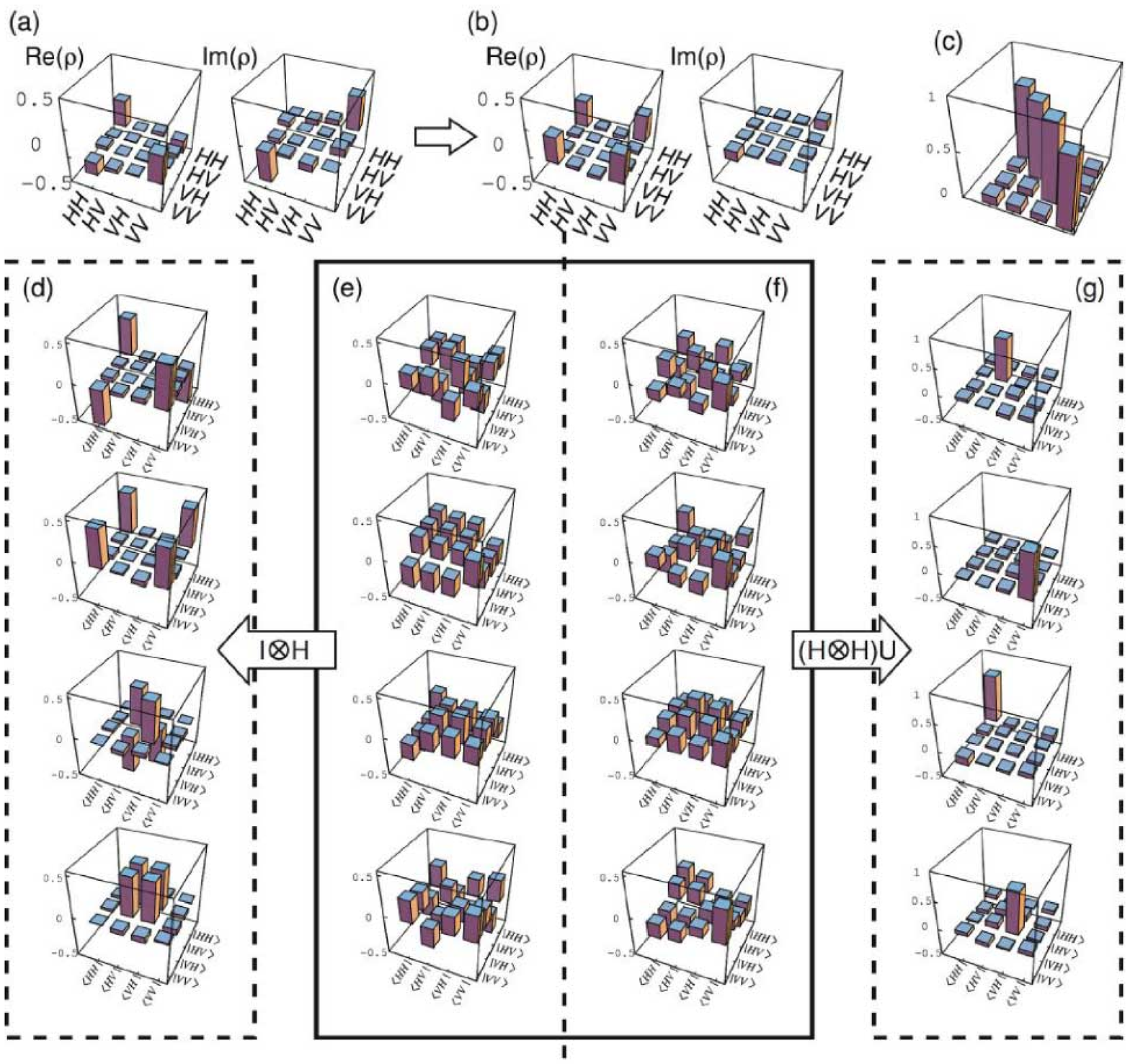}
\vspace{-0.5cm}
\caption{The \textsc{cz} gate operating as a Bell-state analyser. (a) The two qubit entangled state at the output of the fibres and (b) transformed to the $\phi^{+}$ Bell-state. (c) The measured truth table: the average probability of success is $0.78\pm0.03$. (d-g) Transformation of near-maximally entangled states to near-separable states by a \textsc{cz} gate Bell-state analyser. (d) The input Bell-states determined from (e) the measured input states with the second qubit rotated by a Hadamard. (f) The measured output states, (g) transformed by applying local rotations to each qubit (see text).}
\label{bell}
\end{center}
\vspace{-0.9cm}
\end{figure}

We further test the \textsc{cz} gate by operating it as a Bell-state analyser of the entangled continuous-wave input states \cite{kw-pra-60-773}. Due to the geometry of the source, and birefringence and geometric effects in the single mode fibers, the near-maximally entangled state produced is of the form $|HH\rangle+e^{i\varphi}|VV\rangle$.  We use quantum state tomography \cite{White,ja-pra-64-052312} to characterize the source state [Fig.~\ref{bell}(a)]. The tangle $T=0.93 \pm 0.01$ and linear entropy $S_{L}=0.05 \pm 0.01$ show that this state is highly entangled and highly pure, the fidelity with a maximally entangled state is $F=98.0 \pm 0.4$\%. We determine that $\varphi$=-2.094 radians, and by using the input quarter- and half-wave plates [Fig.~\ref{schematic}(b)] to perform appropriate single qubit unitaries on each qubit we can transform the state of Fig.~\ref{bell}(a) to any desired maximally entangled state of linear polarisation. Figure \ref{bell}(b) shows the example where we have produced the state $|HH\rangle+|VV\rangle$ with fidelity $F_{\phi^{+}}=96.1 \pm 0.2$\%; $T=0.96 \pm 0.01$ and $S_{L}=0.02 \pm 0.01$.

To quantify the performance of the \textsc{cz} gate as a Bell-state analyser, we produced the four maximally entangled states:
$|\psi'^{\pm}\rangle  =  |HA\rangle\pm|VD\rangle$;
$|\phi'^{\pm}\rangle  =  |HD\rangle\pm|VA\rangle$, 
where $D\equiv(|H\rangle+|V\rangle)/\sqrt{2}$ and $A\equiv(|H\rangle-|V\rangle)/\sqrt{2}$. These are just the usual four Bell states, with the second qubit rotated by a Hadamard so that they can be discriminated by the \textsc{cz} gate. The four experimentally produced density matrices are shown in Fig.~\ref{bell}(e): the average of their fidelities is $\bar{F}=95.8\pm0.7$\%; the average of the tangles and linear entropies are $\bar{T}=0.94\pm0.02$ and $\bar{S}_{L}=0.04\pm0.01$, respectively. For ease of visualisation, we have numerically rotated these states into the more familiar form by applying a Hadamard gate to the second qubit [Fig.~\ref{bell}(d)]. 

An ideal \textsc{cz} gate would take the four maximally entangled states $|\psi'^{\pm}\rangle$, $|\phi'^{\pm}\rangle$ to the four separable orthogonal states: $|DD\rangle$, $|AD\rangle$, $|DA\rangle$, and $|AA\rangle$ respectively. For the four input states in Fig.~\ref{bell}(e), the measured output density matrices are shown in Fig.~\ref{bell}(f). In fact, they are close to the four orthogonal separable states $(|H\rangle\pm e^{i\varphi_{1}}|V\rangle)\otimes(|H\rangle\pm e^{i\varphi_{2}}|V\rangle)$, where $\varphi_{1}=-3.07$ and $\varphi_{2}=0.32$ as determined by a best fit. For ease of visualisation, we have rotated these states into the logical basis in Fig.~\ref{bell}(g). The average of the fidelities between all combinations of the measured output states is $24\pm5$\% (ideally zero), demonstrating that the states are close to orthogonal. Their average tangle, $\bar{T}=0.04\pm0.05$, and linear entropy, $\bar{S}_{L}=0.42\pm0.07$, indicates that they are unentangled, albeit somewhat mixed. This circuit is working quite well as a Bell-state analyser.

The average fidelity of the measured output states with the above separable states is $F=79\pm3$\%: if we analyzed the output of the circuit in this rotated basis, we would correctly identify the Bell-state with a probability of 79\%. More directly, we can measure each of the separable states for each Bell-state input by explicitly analysing in the rotated basis, which gives the directly measured truth table for the \text{CZ} gate when operated as a Bell-state analyser. The results are shown in Fig.~\ref{bell}(c), and the average probability of success is $78\pm3$\%, in agreement with the tomography results.

It is interesting to note that whenever a postselected event occurs, the Bell measurement has effectively discriminated one of four input wave plate settings applied to a single input qubit. That is to say, two bits of classical information (representing the four wave plate settings) have been encoded into a single qubit. This is reminiscent of quantum dense coding \cite{be-prl-69-2881,sc-prl-93-040505,ma-prl-76-4656}, although because the Bell measurement is non-deterministic, a protocol using this gate would be less efficient than ordinary classical communication.  Nevertheless, this still demonstrates the power of entanglement for dense coding given a deterministic Bell analyser, such as can be constructed in principle using measurement-induced nonlinearity.

In summary, we have proposed and demonstrated a new architecture for entangling optical gates. The key advantage of this new gate architecture is its simplicity and suitability for scaling---it requires only one nonclassical mode matching condition, and no classical interferometers. This is very promising for micro-optic and integrated-optic realisations of this gate, where extremely good mode matching can be expected. Finally, we have demonstrated the operation of this gate as a Bell-state analyser which has the advantage of higher success probability and no ancillae compared to alternative recent demonstrations \cite{ZhaoPRL,wa-zeil-PRA-2005}.

This work was supported by the Australian Research Council (ARC), the Queensland Government, and the US Advanced Research and Development Agency (ARDA). RP acknowledges support from the Austrian Science Foundation (FWF). We wish to acknowledge Rohan Dalton for valuable discussions.


\begin{thebibliography}{19}
\bibitem{note2} The more familiar \textsc{cnot} gate is formed by applying a Hadamard gate \textsc{h} to the target qubit before and after a \textsc{cz} gate.
\bibitem{PPBSnote} The PPBSs  are cube beam splitters with appropriately specified multilayered
dielectric stacks purchased from Asahi Spectra (702.2nm) and Special Optics (820nm).
\bibitem{note} This involves inputting identical ensembles of 16 separable states into the circuit and performing a set of 36 measurements for each---36 measurements form an overcomplete set which increases accuracy.
\bibitem{MCnote} The errors were estimated by doing a 1,000 run Monte-Carlo simulation of the whole process tomography analysis, with Poissonian noise added to the count statistics in each run.
\end{thebibliography}
\end{document}